\documentclass[floatfix,twocolumn,aps,prl,showpacs]{revtex4}
\usepackage{dcolumn}
\usepackage{color}
\usepackage{graphics}
\usepackage{epsfig}

\begin{document}

\title{Tight-binding study of structure and vibrations of amorphous silicon}

\author{J.~L. Feldman}
\affiliation{Center for Computational Materials Science, Naval
        Research Laboratory, Washington, DC 20375}

\author{N. Bernstein}
\affiliation{Center for Computational Materials Science, Naval
        Research Laboratory, Washington, DC 20375}

\author{D.~A. Papaconstantopoulos}
\affiliation{Center for Computational Materials Science, Naval
        Research Laboratory, Washington, DC 20375}

\author{M.~J. Mehl}
\affiliation{Center for Computational Materials Science, Naval
        Research Laboratory, Washington, DC 20375}

\date{\today}

\begin{abstract}

We present a tight-binding calculation that, for the first time,
accurately describes the structural, vibrational and elastic properties
of amorphous silicon. We compute the interatomic force constants
and find an unphysical feature of the Stillinger-Weber empirical
potential that correlates with a much noted error in the radial
distribution function associated with that potential.  We also find
that the intrinsic first peak of the radial distribution function
is asymmetric, contrary to usual assumptions made in the analysis of
diffraction data. We use our results for the normal mode frequencies
and polarization vectors to obtain the zero-point broadening effect
on the radial distribution function, enabling us to directly compare
theory and a high resolution x-ray diffraction experiment.

\end{abstract}


\maketitle

Amorphous silicon (a-Si) is a prototype for continuous-random-network
covalent glasses that, with some hydrogen content, has
technological applications as a relatively inexpensive electronic
material.  While the basic structure of a-Si is believed to
be a four-fold-coordinated continuous random network, detailed
information about network connectivity and defects is lacking.
Atomic resolution structure is very difficult to determine
directly, and experiments have relied on unusual or indirect
probes such as variance coherence microscopy~\cite{treac} and Raman
spectroscopy~\cite{beem,vink1} as well as on more standard techniques
such as diffraction~\cite{fort,laaz} and EXAFS~\cite{wakagi,new_exafs}.
The experimental measurements suggest significant deviation
from a continuous random network, including average coordination
that is significantly less than 4 (e.g. Ref.~\onlinecite{laaz})
and that unannealed samples may be paracrystalline~\cite{treac}.
Many empirical-potential simulations have been done, but it is not
clear if empirical potentials are accurate enough to give reliable
results for properties, such as coordination defects, that depend
on bond breaking and bond formation.  A number of simulations of
a-Si structure have used electronic-structure based methods, which
are generally among the most reliable for solid state systems
(e.g. Refs.~\onlinecite{stich,pbis1,durdr,klein}).  However,
none have carefully compared the radial distribution function
(RDF) to high resolution experiments~\cite{laaz}, and none included
quantum-mechanical vibrational effects.  Another important question
concerns the vibrational properties of a-Si, which give us information
about the structure and the interactions of atoms in the material.  The
vibrational density of states (VDOS) was measured experimentally using
inelastic neutron scattering (INS)~\cite{kama}.  Empirical-potential
simulations have been used to analyze vibrational properties in
detail~\cite{allen}, but all show significant errors in the shape
of the VDOS or in other properties.  While the VDOS of a-Si has been
simulated with electronic structure methods~\cite{stich,pbis,nakdra},
the underlying force constants themselves have not been analyzed.
There have been many studies of force constants in crystalline Si,
which shows unusual phonon dispersion and force constants that
oscillate in magnitude as a function of distance~\cite{kane,rign}.

We study the elastic constants, vibrational properties, and structure
of a-Si using a tight-binding (TB) total-energy method.  We find
elastic constants and VDOS that are in good agreement with experiment,
and qualitatively better than empirical-potential simulations.
The structure has a sharp first-neighbor RDF peak that agrees very well
with experiment when zero-point and thermal broadening is included.
This peak is significantly non-Gaussian, calling into question the
coordination-statistics analysis of previous diffraction experiments.

We use the Naval Research Laboratory (NRL) TB method~\cite{rec,brn}.
The non-orthogonal $sp^3$-basis TB model has been shown to accurately
describe the elastic constants and phonon dispersion in crystalline
Si and the electronic density of states for a highly defected
amorphous model~\cite{brn}.  To generate the a-Si models we relax
using TB-calculated forces a-Si models generated by other techniques.
The NRL-TB model is used to calculate the energy of the structure
and the atomic forces~\cite{kirch}.  The conjugate-gradient method is
applied to find mechanical-equilibrium positions at a fixed volume,
employing the criterion that components of atomic forces be less
than $10^{-3}$~eV/{\AA}.  The relaxation procedure is carried out at
several volumes to obtain results at zero pressure, but components
of the stress tensor, generally of magnitude less than 0.8~GPa, remain.

One model, which we denote TB1, is generated by relaxing (using
TB) a 216 atom perfect continuous-random-network model~\cite{woot}
with periodic boundary conditions relaxed with a Keating interatomic
potential~\cite{www}.  The TB-relaxed model is perfectly four-fold
coordinated, with 1.3\% lower density than the crystal, compared
to 1.7\% lower density measured experimentally~\cite{laaz}.
The bond-angle distribution has a RMS deviation of 11$^\circ$
from the average value of 109.2$^\circ$, in close agreement with
relaxed {\it ab initio} calculation~\cite{durdr} and analysis
of experiment~\cite{fort}.  A second model, which we denote TB2,
is generated by relaxing a structure from a molecular-dynamics
simulation of the rapid quenching of liquid Si using the environment
dependent interatomic potential~\cite{edip}.  The TB2 structure is
slightly more dense than TB1, but still about 0.5\% less dense than
the crystal.  The energy is 28~meV/atom lower than the TB1 energy,
despite the presence of 6\% 5-fold and 0.46\% 3-fold coordinated atoms
(corresponding to an average coordination of 4.05)~\cite{model_energy}.
The RMS bond-angle deviation is 12.5$^\circ$, although the
distribution has wide, non-Gaussian wings;  excluding 2\% of the
bond-angles reduces the RMS deviation to 10.4$^\circ$.  We also
show some comparisons with results using the Stillinger-Weber (SW)
interatomic potential~\cite{sw}.  The SW potential, which includes
radial and bond-angle terms, is one of the most often used potentials
for simulations of Si.  We use a structure (Ref.~\onlinecite{fbr},
Table~II, model~IV) generated by relaxing with SW the same starting
structure as TB1.  Finally, we note that while it is possible to use
electronic structure methods to generate amorphous structures from
procedures that are less dependent on the initial structure, it is
very expensive computationally.  The difficulty in fully annealing
the structure seems to lead to a consistent overestimate of the width
of the first-neighbor peak in the RDF~\cite{stich,pbis1}.

\begin{table}
\caption{Selected elastic constants $c$, bulk modulus $B$ and Young's
modulus $E$ ($10^{11}$~dyn/cm$^2$).  The index $i$ varies from 1 to 3,
and $j$ from 4 to 6.}
\label{tab1}

\begin{tabular}{lcccc}
 \hline
 		& TB1		& TB2	& Exp./FP			& SW$^{(a)}$\\ \hline
c$_{ii}$	& 16.31-16.45	& 15.06-16.00	&  13.8$^{(b)}$,17(2)$^{(c)}$	& 11.94-13.11 \\
c$_{jj}$	& 5.68-5.84	& 5.26-5.56	&  4.8$^{(b)}$, 4.5$^{(a)}$	& 2.54-3.21 \\
c$_p^{(d)}$	& 5.77		& 5.06	&  "				& 2.62\\
c$_{12}$	& 4.77		& 5.32	&  				& 6.69\\
B		& 8.73		& 8.99	&  5.9$^{(e)}$,8.25$^{(f)}$	& 8.52\\
E		& 14$^{(g)}$		& 13$^{(g)}$	&  12.4(3)$^{(a)}$		& 7$^{(g)}$\\
   		&		& 	& 11.7(5)-13.4(5)$^{(h)}$	& \\ \hline
\multicolumn{5}{l}~(a) Ref. \onlinecite{fbr};~(b) Ref. \onlinecite{zhang};~(c) Ref. \onlinecite{grim}; \hfill\\
\multicolumn{5}{l}~(d) Defined here as (c$_{11}$-c$_{12}$)/2;~ (e) Ref. \onlinecite{tanaka} \hfill\\
\multicolumn{5}{l}~(f) Ref. \onlinecite{durdr};~(g) based on values of c$_{12}$ and c$_p$;\hfill\\
\multicolumn{5}{l}~(h) Ref. \onlinecite{kusch}. \hfill
\end{tabular} 
\end{table}

The relaxed static lattice TB elastic constants $c_{ij}$ were
obtained by the method of homogeneous deformation.  The TB
results~\cite{elastic_isotropy} are compared in Table~\ref{tab1}
with results of first-principles (FP)~\cite{durdr} calculations, SW
calculations, and several experiments on dense samples (a wider range
of shear values are quoted in Table~V of Ref.~\onlinecite{pliu}).
Although there is some deviation between the two TB structures it
is small.  While ultrasonic measurements of elastic properties are
not available for a-Si, the Young's modulus $E$ can be measured
with a vibrating reed apparatus, and other elastic constants can be
inferred from spectroscopic studies.  Our TB results for both models
are close to the experimental values, although our value of $c_{44}$
is likely 10--20\% too large.  The SW empirical potential results
are significantly worse in comparison with experiment.


\begin{figure}    

\centerline{\resizebox{\columnwidth}{!}{\includegraphics{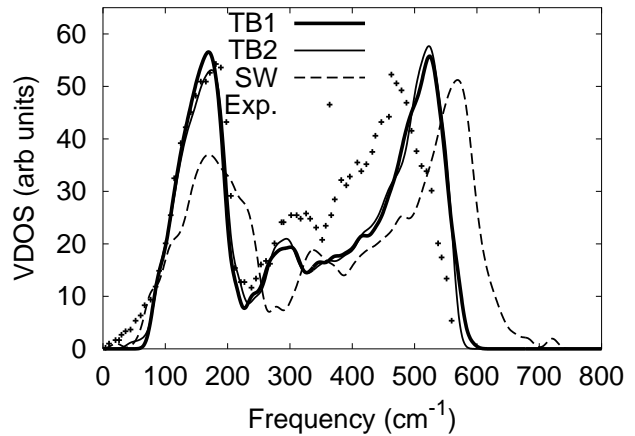}}}
    \caption{Vibrational density of states of a-Si. A Gaussian broadening
of $FWHM = 20$~cm$^{-1}$ is employed.
    The experimental data are from Ref.~\cite{kama}.}

    \label{gom}

\end{figure}


The VDOS is calculated from a dynamical matrix approach.  The matrix
elements $\Phi_{\alpha\beta}(i,j) \equiv \Delta F_\alpha(i)/\Delta
u_\beta(j)$ are calculated using the TB forces with a central-finite
difference approach that eliminates all odd-order anharmonic terms
in the potential energy~\cite{explain}.  The TB VDOS for structures
TB1 and TB2 are compared with SW results and INS~\cite{kama} measurements in Fig.~\ref{gom}.  For both
structures the TB calculation yields the overall shape very well;
it exactly describes the low frequency TA peak, gives a slightly too
small frequency of the LA peak (300 $cm^{-1}$) and about a 10$\%$
percent too high frequency of the high frequency TO peak.  The TB
results are a qualitative improvement over results based on the SW
potential, as shown in the figure, and they are in good agreement with
{\it ab initio} results for a 216 atom structural model~\cite{nakdra}.


\begin{figure}

\centerline{\resizebox{\columnwidth}{!}{\includegraphics{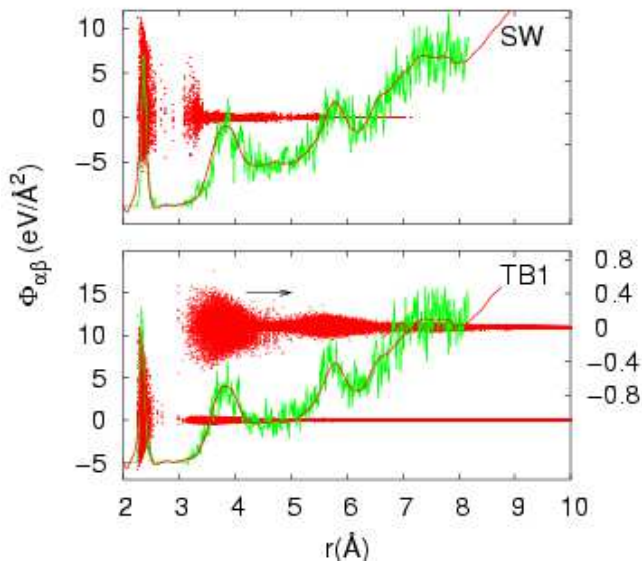}}}
    \caption{Force constants between pairs of atoms in SW (top) and
    TB1 (bottom) relaxed structures (dots).  Superimposed are the
    corresponding $J(r)$ functions (jagged lines) in arbitrary units
    and the experimental, annealed-sample, results of Ref.~\cite{laaz}
    for $J(r)$ (smooth lines).  The upper scatterplot in the TB panel
    is a magnification of the smaller magnitude force constants.}

\label{fcs}
\end{figure}

The range of the effective interactions in the solid can give us
information about the physics of the interactions, and can guide
the development of approximations such as empirical potentials.
In Fig.~\ref{fcs} we plot all of the cartesian force constants
between pairs of atoms with interatomic distances less than 10~{\AA}.
The difference in range between the SW results and the TB results
is easy to see: The SW interactions are large up to about 3.5~{\AA},
and go to exactly zero at twice the SW cutoff of 3.75~{\AA}.  The TB
interactions are already quite small at 2.8~{\AA}, but do not go
to zero even at 10~{\AA}.  This comparison of TB and SW leads to a
view of interactions in the solid that is more subtle than the usual
assumption that empirical potentials are short ranged and that the
real interactions are long ranged: The SW potential interactions go to
zero at a range that is too short, but at intermediate distances the
interactions are too strong.  We also note that the preponderance of
force constants as a function of interatomic distance give a clear
envelope function that has an oscillatory behavior which matches
the RDF peak positions. This is qualitatively similar to the case of
the crystal, even though the explanations for the oscillation in the
crystal do not apply to the amorphous structure~\cite{kane,rign}.

The problem with the SW potential is a direct consequence of
the form of the potential.  In the amorphous there are pairs of
atoms in the second-neighbor peak with distances smaller than
the SW cutoff.  It is clear from the TB force constants that the
effective interactions for these pairs is qualitatively different
from the first-neighbor interactions.  However, in the SW simulation
these second-neighbor pairs interact through terms that are meant to
describe the interactions of first-neighbor atoms.  In particular, the
two-body contribution has strong negative curvature at these distances,
and the three-body terms include contributions from triplets with
a vertex angle that does not correspond to an atom with two $sp^3$
orbitals in bonding configurations.  These two types of contributions
lead to the unphysically large force constants in the SW results at
this range of distances.   The range of incorrect force constants
also coincides with the shoulder in the SW RDF that is not observed
in our TB results or in the experimental measurements~\cite{vink}.


\begin{figure}

\centerline{\resizebox{\columnwidth}{!}{\includegraphics{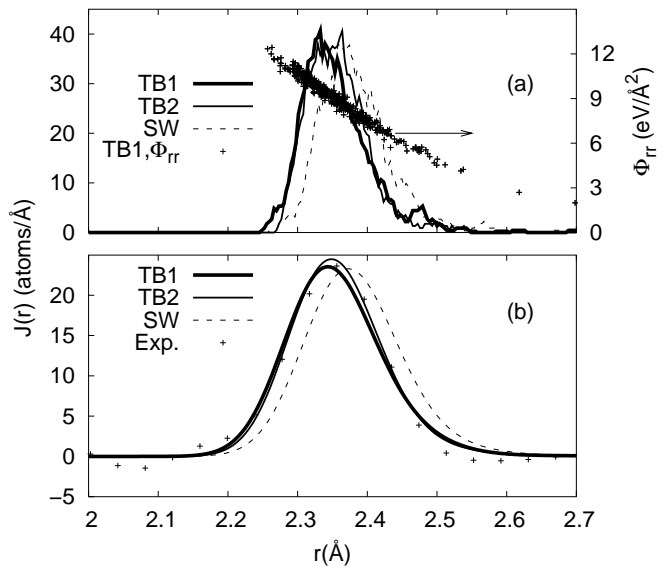}}} 

    \caption{(a) First peak of the static RDF
    and TB1 bond stretching force constants.  (b) Broadened results
    corresponding to T=10K in comparison with experiment
   (annealed sample).\cite{laaz}}

\label{rdf}
\end{figure}

The distribution of force constants gives us information about the
types of effective interactions between bonded atoms.  Under the
first peak of the RDF the largest positive cartesian force constants
are twice the magnitude of the largest negative force constants
for both SW and TB.  This relation is consistent with an effective
bond-stretching interaction for first-neighbors.  We plot the results
for the bond-stretching components in a plot as a function of $r$
(Fig.~\ref{rdf}a). The radial force constants decrease with increasing
$r$ as one expects from a physically reasonable first-neighbor bonding
potential.  Pairs with large (small) interatomic bond stretching force
constants will have small (large) relative mean square displacements,
so these results clearly have an impact on the nature of the broadening
of the RDF.

Very little attention has been given in the literature to the shape of
the first peak in the RDF $J(r)$~\cite{herrero}.  This peak has been
measured very carefully at $T=10$~K with x-ray diffraction, using high
energy photons and high resolution, i.e., large $Q_\mathrm{max}$, by
Laaziri {\it et al.}~\cite{laaz}.  They obtain a fit of their data to
a Gaussian, with average coordination of 3.88${\pm .01}$ (3.79${\pm
.01}$) for the annealed (unannealed) sample. In Fig.~\ref{rdf}a we
plot the first peak of the static $J(r)$ for models TB1 and TB2, and
the SW results.  The TB static peak is asymmetric, and its width is
significantly larger than the static-disorder estimate by Laaziri {\it
et al.} In order to compare directly with the experimental $J(r)$ it
is necessary to properly take account of the zero-point and thermal
broadening.  The quantity measured by the x-ray experiment is, in
the small-displacement limit,
$$J(r)=\frac{1}{N}\sum_{i,j=1}^N\frac{1}{\sqrt{2{\pi}U^r_{ij}}}exp(-(r_{ij}-r)^2/(2U^r_{ij})),$$
where U$^r_{ij}\equiv\langle({\bf \hat{r}}_{ij}\cdot {\bf
u}_{ij}$)$^2\rangle$. Thus we need the mean-squared relative
displacements, for pairs of atoms, along the interatomic vector
direction.  We calculate them within the harmonic approximation
at $T=10$~K using our computed vibrational modes.  Since $T=10$~K
essentially corresponds to $T=0$~K for these considerations, what
we obtain is the minimum measurable width for the first peak in the
RDF of amorphous silicon.  As seen in Fig.~\ref{rdf}b the results
are in agreement with experiment, aside from a small skewing of the
theoretical function to large $r$.  Although it has not been observed
in a-Si, this type of asymmetry has been observed in EXAFS of amorphous
germanium~\cite{ge_exafs}.  Both the TB1 and TB2 models, despite the
very different originating structure and differences in coordination
defects, show nearly identical RDF first peaks.  The good agreement
with experiment of the broadened RDF suggests that our static peak
width is correct, and that Laaziri {\it et al.} underestimate the
static disorder contribution to the broadening.  This may be caused
by inaccuracy in the polycrystalline $J(r)$ that is used to estimate
the dynamic broadening.  In the experiments a lower $Q_\mathrm{max}$
(35~\AA$^{-1}$) was used for the polycrystal than for the amorphous
structure (55~\AA$^{-1}$), although the former is expected to have
a narrower first peak.  Numerous other treatments using EXAFS or
diffraction have not been considered here because they all use too
low values of $Q_\mathrm{max}$ for obtaining reliable information on
the first peak.  The only other theoretical study of quantum effects
in $J(r)$ is by Herrero~\cite{herrero}, who used the SW potential but treated the
quantum-effects on the nuclear vibrations exactly.  Our results using
the SW potential are presented in Fig.~\ref{rdf}.  The result for the
amount of zero-point broadening is consistent with Herrero's work,
although due to differing approximations a direct comparison is not
possible.  We note that the Wooten model on which both the SW and
TB1 models are based yields a {\em static} $J(r)$ (not shown) that
is quite symmetric, and as broad as the {\em experimental} breadth.

To conclude, we have shown that the NRL-TB method can reliably
compute structural, vibrational, and elastic properties of a-Si.
The results are nearly identical for two structural models, one with
perfect four-fold coordination and one with several atomic percent
coordination defects.  We have presented the first discussion of
force constants in a-Si, which has revealed limitations of the most
frequently used empirical potential for silicon.  Our calculated
elastic constants fall within the range of experimental values
for imperfect samples prepared under various conditions.  We have
also carefully studied the first peak in the radial distribution
function. We observe a clear asymmetric peak in the case of the static
quantity which is not observable experimentally. We have included the
(essentially) zero-point broadening effects in $J(r)$ to obtain the
experimentally measured quantity.  Our two structural models, which
have average coordinations of 4.00 and 4.05, respectively, reproduce
the first peak in the experimental $J(r)$ (for the annealed sample)
except for a slight asymmetry still present in the broadened result.
We believe that such an asymmetry is expected on physical grounds
and that perhaps it has been ``missed'' experimentally because of the
challenging analysis required to obtain $J(r)$ from the diffraction
results.

This work was supported by the U.S. Office of Naval Research.
We are also grateful to Dr. S.~Roorda for a helpful communication
and for sending us the x-ray data of Ref.~\cite{laaz} for the radial
distribution function. We thank Dr. S. Richardson for a helpful
conversation.



\begin{thebibliography}{99}

\bibitem{treac} M.M.J. Treacy, J.M. Gibson, and P.J. Keblinski, J. of Non-Cryst. Solids,
{\bf 231}, 99 (1998).

\bibitem{beem} D. Beeman, R. Tsu, and M.F. Thorpe, Phys. Rev. B {\bf 32}, 874 (1985). 

\bibitem{vink1}  R.L.C. Vink, G.T. Barkema, and W.F. van der Weg, Phys. Rev. B {\bf 63}, 115210 (2001) and references therein.

\bibitem{fort} J. Fortner and J.S. Lannin, Phys. Rev. B {\bf 39}, 5527 (1989).

\bibitem{laaz} K. Laaziri, S. Kycia, S. Roorda, M. Chicoine, J.L. Robertson, J. Wang, and S.C. Moss,
Phys. Rev. Lett., {\bf 82}, 3460 (1999) and references therein;
ibid., Phys. Rev. B {\bf 60}, 13520 (1999).

\bibitem{wakagi} M.Wakagi, K. Ogata, and A. Nakano, Phys. Rev. B {\bf 50},
10666 (1994-I); A. Filipponi, F. Evangelisti, M. Benfatto, S. Mobilio, and
C.R. Natoli, Phys. Rev. B {\bf 40}, 9636 (1989).

\bibitem{new_exafs} C.J. Glover, G.J. Foran, and M.C. Ridgway, Nuc. Instr.
and Methods in Physics Res. B {\bf 199}, 195 (2003).

\bibitem{stich} E.g. I. Stich, R. Car, and M. Parrinello, Phys. Rev. B {\bf 44}, 11092 (1991-II)

\bibitem{pbis1} P. Biswas, Phys. Lett. A {\bf 282}, 294 (2001).

\bibitem{durdr} M. Durandurdu and D.A. Drabold, Phys. Rev. B {\bf 64}, 014101 (2001).

\bibitem{klein} P. Klein, H.~M. Urbassek, and T. Frauenheim, Comp. Mat. Sci. {\bf 13},
    252 (1999).

\bibitem{allen} P.B. Allen, J.L. Feldman, J. Fabian, and F. Wooten, Phil.
Mag. B {\bf 79}, 1715 (1999).

\bibitem{pbis} P. Biswas, Phys. Rev. B {\bf 65}, 125208 (2002).

\bibitem{nakdra} S.M. Nakhmanson and D.A. Drabold, J. of Non-Cryst. Solids, {\bf 266-269}, 156 (2000).

\bibitem{kane} E.O. Kane, Phys. Rev. B {\bf 31}, 7865 (1985).

\bibitem{rign} G.-M. Rignanese, J.-P. Michenaud, and X. Gonze, Phys. Rev. B {\bf 53}, 4488 (1996).

\bibitem {rec} R.E. Cohen, M.J. Mehl, and D.A. Papaconstantopoulos, Phys. Rev. B {\bf 50}, 14695 (1994); M.J. Mehl and D.A. Papaconstantopoulos, ibid., {\bf 54} 4519 (1996).

\bibitem{brn} N. Bernstein, M. J. Mehl, D. A. Papaconstantopoulos, N. I. Papanicolaou, M. Z. Bazant,and E. Kaxiras, Phys. Rev. B {\bf 62},4477 (2000); ibid., {\bf 65}, 249902 (2002) (E).

\bibitem{woot} F. Wooten, private communication.

\bibitem{www} F. Wooten, K. Winer, and D. Weaire, Phys. Rev. Lett. {\bf 54},1392 (1985).

\bibitem{kirch} F. Kirchhoff, M.J. Mehl, N.I. Papanicolaou, D.A. Papaconstantopoulos, and F.S. Khan, Phys. Rev. B {\bf 63}, 195101 (2001).

\bibitem{edip} M. Bazant, private communication; see also J.F.
Justo, M.Z. Bazant, E. Kaxiras, V.V. Bulatov, and S Yip, Phys.
Rev. B {\bf 58}, 2539 (1998).

\bibitem{model_energy} Although this energy difference suggests the
possibility of energetically favorable coordination defects, the small
system sizes and minimal annealing make this conclusion uncertain.

\bibitem{fbr} J.L. Feldman, J.Q. Broughton, and F. Wooten, Phys. Rev. B {\bf 43}, 2152 (1991) and references therein.

\bibitem{zhang} X. Zhang, J.D. Comins, A.G. Every, P.R. Stoddart, W. Pang, and T.E. Derry, Phys. Rev. B {\bf 58},
13677 (1998). 

\bibitem{grim} M. Grimsditch, W. Senn, G. Winterling, and M.H. Brodsky, Sol. State Commun. {\bf 26}, 229 (1978).

\bibitem{tanaka} K. Tanaka, Sol. State Commun. {\bf 60}, 295 (1986).

\bibitem{kusch} Kuschnereit et al., Applied Phys. A (Mat. Sci. and Proc.) {\bf 61}, 269 (1995).

\bibitem{sw} F.H. Stillinger and T.A. Weber, Phys. Rev. B {\bf 31}, 5262 (1985). 

\bibitem{pliu} R.O Pohl, X. Liu, and E. Thompson, Rev. of Mod. Phys. {\bf 74}, 991 (2002).

\bibitem{elastic_isotropy} Due to finite-size effects, our results do
not exactly satisfy the expected isotropy conditions on the elastic
constants of an amorphous material.

\bibitem{kama} W.A. Kamitakahara, C.M. Soukoulis, H.R. Shanks, U. Bucheneau, and G.S. Grest, Phys. Rev. B {\bf 36}, 6539 (1987).


\bibitem{explain} We tacitly assume that the cell size is big enough
that the force on an atom due to the displacement of a different atom
and that of its periodic images can be ascribed solely to the closest
displaced atom.

\bibitem{vink} R.L.C. Vink, G.T. Barkema, W.F. van der Weg, and N. Mousseau, J. Non-Cryst. Solids, {\bf 282}, 248 (2001).

\bibitem{herrero}  C. P. Herrero, Europhys. Lett. {\bf 44}, 734 (1998);
    C. P. Herrero, J. of Non-Cryst. Solids {\bf 271}, 18 (2000).

\bibitem{ge_exafs} M.C. Ridgway, C.J. Glover, K.M. Yu, G.J. Foran, C. Clerc,
J.L. Hansen, and A.N. Larsen, Phys. Rev. B {\bf 61}, 12586 (2000).

\end{thebibliography}
\end{document}